\documentclass[a4paper,12pt]{article}
\usepackage[utf8]{inputenc}
\usepackage{amsmath}
\usepackage{float}
\usepackage{graphicx}
\usepackage{setspace}
\usepackage[numbers]{natbib}
\usepackage{hyperref}
\usepackage{tikz}
\usetikzlibrary{shapes,arrows}

\makeatother

\begin{document}

\begin{titlepage}
\vspace*{-25mm}
\begin{flushright}
%SHEP-14-00\\
\end{flushright}
%\vspace*{5mm}

\begin{center}
{ \sffamily \Large \bf
Fine tuning in low and large $\tan \beta$ regions in the cE$_6$SSM
} 
\\[8mm]
Maien~Y.~Binjonaid$^{a,b}$
\footnote{E-mail: \texttt{mymb1a09@soton.ac.uk, maien@ksu.edu.sa}}
%\qquad\qquad\qquad\qquad\qquad\qquad\qquad
\\[3mm]
{\small\it
$^a$ School of Physics and Astronomy, University of Southampton,\\
Southampton, SO17 1BJ, U.K.\\[2mm]
$^b$ Department of Physics and Astronomy, King Saud University,\\
Riyadh 11451, P. O. Box 2455, Saudi Arabia
}\\[1mm]
\end{center}
\vspace*{0.5cm}

\begin{abstract}

The Electroweak sector in $E_6$ supersymmetric models is subject
to a degree of fine tuning in the percent to permil level. This can be attributed to
the experimental limits on both the mass of the $Z'$ boson associated
with the extra $U(1)'$ symmetry in the model, as well as the masses
of naturalness-related sparticles (which is a general source of tuning
in supersymmetric models). The degree of tuning can be
smaller than that in the minimal supersymmetric standard model with
universal fundamental parameters (the constrained MSSM). We show this
by quantifying the fine tuning in regions of the parameter space of
the constrained exceptional supersymmetric standard model (cE$_6$SSM)
corresponding to values of $\tan \beta$ below and above 10. It is
found that, a Higgs mass $m_h \sim 125$ GeV, a gluino mass $m_{\tilde{g}} \sim 1.5$ TeV,
and a $Z'$ boson mass $m_{Z'} \sim 3.8$ TeV correspond to fine tuning
in the $0.2\% \ (0.1\%)$ level for $\tan \beta = 30 \ (5)$. 
\end{abstract}

\end{titlepage}

\section{Introduction}

Naturalness, which can be understood as the requirement that observable
quantities in a given model does not possess large and unexplained
fine tuning (see \citep{'tHooft:1979bh}), has been a leading principal
for developing theories beyond the Standard Model (SM). The quadratic sensitivity
of the Higgs mass-squared parameter ($m_H^2$) to the scale of new
physics, be it the Planck scale at $10^{19}$ GeV or a scale at which
heavy masses may exist (e.g. the Grand Unification (GUT) scale $M_{GUT}=10^{16}$ GeV),
has led the community to suggesting the existence of new physics at
low scale near 1 TeV (see \citep{PhysRevD.13.3333,Susskind:1978ms}).
This is due to the fact that in the absence of new physics at the
low scale, the parameter $m_H^2$, which is proportional to the measured
value of the vacuum expectation value (VEV) of the Higgs field ($v=246.22$ GeV),
will need to be carefully fine tuned, order by order in perturbation
theory, against the cutoff of the new scale (or any heavy mass), thereby
destabilizing the Electroweak scale. For example, if the scale of
such masses is the GUT scale ($\sim 10^{16}$ GeV), then the degree
of tuning is roughly 1 part in $10^{32}$. 

Among the well-motivated and most studied theories beyond the standard
model that might appear at the low scale are supersymmetric (SUSY)
models, in which $m_H^2$ is protected from large radiative corrections
by the symmetry (for a pedagogical review see \citep{Martin:1997ns}).
However, the LHC is pushing the scale of SUSY close to or above 1
TeV \citep{ATLAS-CONF-2013-053,ATLAS-CONF-2013-062}, thereby placing
the concept of Naturalness to the experimental test. 

SUSY models differ in their predictions of observables, such as the
mass of the Higgs and the Z bosons ($m_Z$). While some models require
large contributions from radiative corrections as in the MSSM, other
models can accommodate a 126 GeV Higgs even at tree-level. The E$_6$SSM
is such a model \citep{King:2005jy,King:2005my} (introduced in Section~\ref{sec:2}).
However, it is a general feature of SUSY models that, the more the
SUSY scale is pushed up by experiments (i.e. more separation from
the weak scale), the more fine tuning is required in order to correctly
predict the measured values of observables. Additionally, $E_6$ models
like the E$_6$SSM have a new and distinct source of fine tuning which
is the $Z'$ boson \citep{Hall:2012mx} that was investigated in \citep{Athron:2013ipa}.

One of the implications of this tension between the Electroweak scale
and the SUSY scale (also known as the little hierarchy problem) is
that a given model or a specific point in the parameter space becomes
less attractive from the point of view of Naturalness which is usually
used as a criteria to favour models or points in the parameter space
over others. It is important from model building point of view to
learn the degree of fine tuning within a given model and whether or
not it is possible to find regions in the parameter space that have
low fine tuning and correct predictions for the values of observables.
This, then, can be directly related to testing the predictions of
Naturalness at the LHC.

In this note, we probe regions in the parameter space of the cE$_6$SSM
with unexplored fine tuning and quantify it, thereby complementing
the results in \citep{Athron:2013ipa}.

\section{The exceptional supersymmetric standard model \label{sec:2}}

The E$_6$SSM is based on the exceptional Lie group E$_6$, which
contains SO(10) as a subgroup, of which $SU(5)$ is a subgroup. It
is possible then to decompose the fundamental representation of dimension
27 under $SU(5) \times U(1)'$ as,

\begin{equation}
27\rightarrow\underbrace{10_{1}+\overline{5}_{2}}_{\begin{smallmatrix}\text{Quarks \ensuremath{\&}}\\
\text{Leptons}
\end{smallmatrix}}\ \ +\underbrace{1_{5}}_{\text{Singlets}}+\underbrace{1_{0}}_{\text{R-H Nutrinos}}+\underbrace{\overline{5}_{-3}+5_{-2}}_{\begin{smallmatrix}\text{Higgs doublets \ensuremath{\&}}\\
\text{Exotics}
\end{smallmatrix}}
\end{equation}

One requires three 27s to ensure an anomaly free model. Additionally,
extra non-Higgs superfields denoted $(H',\overline{H}')$ coming from
other incomplete representations denoted $(27',\overline{27}')$ are
added in order to ensure gauge coupling unification. Thus, in the
notation of the SU(5) group the complete matter content in the E$_6$SSM
is,

\begin{equation}
3\underbrace{(\overline{5}+10+1_{0}+1_{5}+\overline{5}+5)}_{27}+\underbrace{(H',\overline{H}')}_{27',\overline{27}'}
\end{equation}

At the GUT scale, where the gauge couplings unify,
the E$_6$ group breaks down to the group structure of the standard
model with an additional $U(1)'$ group,

\begin{equation}
SU(3)_{c}\times SU(2)_{L}\times U(1)_{Y}\times U(1)',
\end{equation}

\noindent which survives to low energy scales ($\sim$1 TeV).

In order to prevent rapid proton decay, flavor-changing neutral currents,
and allowing only third generation Higgs doublets ($\hat{H}_{u,d}$)
and SM singlet ($\hat{S}_3$) to couple to matter superfields,
a number of discrete symmetries is imposed, namely, an approximate
$Z_2^H$, and either a $Z_2^L$ or a $Z_2^B$ which specifies two
distinct models allowing the exotic matter to be either diquarks or
leptoquarks. 

The $Z_2^H$ invariant superpotential reads,
\begin{eqnarray}
W_{\rm E_6SSM} &\approx &
\lambda_i \hat{S}(\hat{H}^d_{i}\hat{H}^u_{i})+\kappa_i \hat{S}(\hat{D}_i\hat{\overline{D}}_i)+
f_{\alpha\beta}\hat{S}_{\alpha}(\hat{H}_d \hat{H}^u_{\beta})+ 
\tilde{f}_{\alpha\beta}\hat{S}_{\alpha}(\hat{H}^d_{\beta}\hat{H}_u) \nonumber\\[2mm]
&&+\dfrac{1}{2}M_{ij}\hat{N}^c_i\hat{N}^c_j+\mu'(\hat{H}'\hat{\overline{H'}})+
h^{E}_{4j}(\hat{H}_d \hat{H}')\hat{e}^c_j+h_{4j}^N (\hat{H}_{u} \hat{H}')\hat{N}_j^c 
\nonumber \\[2mm]
&& + W_{\rm{MSSM}}(\mu=0),
\label{Eq:SupPot}
\end{eqnarray}

\noindent where the indices $\alpha, \beta = 1,2$ and $i = 1,2,3$ denote the generations. $S$ is the SM singlet field, $H_u,$ and $H_d$ are the Higgs doublet fields 
corresponding to the up and down types.  Exotic quarks and the additional non-Higgs fields are denoted by $D$ and $H'$ respectively. 

In order to ensure that only third generation Higgs like fields get
VEVs a certain hierarchy between the Yukawa couplings must
exist. Defining $\lambda \equiv \lambda_3$, we impose

$\kappa_i,\lambda_i \gg
f_{\alpha\beta},\,\tilde{f}_{\alpha\beta},\,h^{E}_{4j},\,h_{4j}^N$. Moreover, we do not impose any unification of
the Yukawa couplings at the GUT scale.

Finally, investigations of the Higgs sector, sparticles mass spectrum, dark
matter, gluino phenomenology, flavor physics, gauge coupling unification,
and F-theory origins can be found in \citep{Athron:2009bs,Athron:2009ue,Athron:2010zz,Athron:2011wu,Athron:2012pw,Athron:2012sq,Hall:2009aj,Hall:2010ix,
Hall:2011zq,Belyaev:2012jz,Belyaev:2012si,Howl:2008xz,Howl:2009ds,King:2007uj,Callaghan:2012rv,Callaghan:2013kaa}

\section{Fine tuning}

In a given model, it is possible to quantify the fine
tuning associated with observables by systematically studying their
sensitivity to fractional variations in the GUT scale fundamental
parameters. To capture and quantify this sensitivity, Ellis et. al.
\citep{Ellis:1986yg} proposed a measure that is widely used in the
literature (e.g. \citep{Barbieri:1987fn,Chankowski:1997zh,Agashe:1997kn,Wright:1998mk,Kane:1998im,BasteroGil:1999gu,Kang:2012sy,
Antusch:2012gv,Perelstein:2012qg}) and can be defined as, 

\begin{equation}
\Delta_{a}=\left|\frac{d\ln M_{Z}}{d\ln a}\right|,\label{eq:ft}
\end{equation}
where $M_Z$ is the mass of the Z boson, which can be expanded in terms of 
a set of fundamental parameters as,

\begin{equation} 
 \frac{M_Z^2}{2} \approx \sum_{i=1}^n F_i z_i a_i^2
 \label{zexpnd}
\end{equation}
where, $a$ denotes the fundamental parameters, $z$ is the coefficient corresponding to each parameter, and is calculated numerically using the renormalisation group equations. $F$ is
some factor, possibly, involving $\tan \beta$. 

\begin{figure}[H]
\hfill{}\includegraphics[scale=0.50]{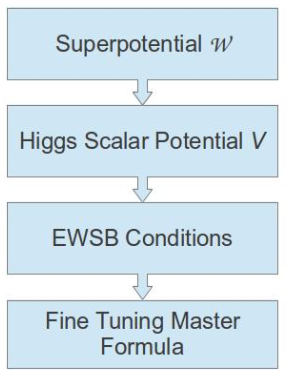}\hfill{}

\caption{Process for deriving the master formula, where EWSB refers to the
Electroweak symmetry breaking conditions obtained by minimizing the
scalar Higgs potential. }

\label{fig:flow}
\end{figure}

Next, using Eq.~\ref{eq:ft} and
following the process sketched in Fig.~\ref{fig:flow}; a master formula
for the fine tuning was derived and presented in \citep{Athron:2013ipa}
where the details of the semi-analytical procedure and code implementations
are provided. 

In this study we scan the parameter space of regions where $\tan \beta$
is either small $\sim 5$ or large $\sim 30$. It is vital to note
that changing $\tan \beta$ affects the running of the renormalisation
group equations which are used to expand low scale parameters in terms
of high scale ones (see Eq.~\ref{zexpnd}), hence allowing the quantifying of fine tuning
in those distinct regions.

\section{Results and discussion}

\begin{figure}[H]
\begin{minipage}[t][1\totalheight][c]{0.45\columnwidth}%
\includegraphics[scale=0.32]{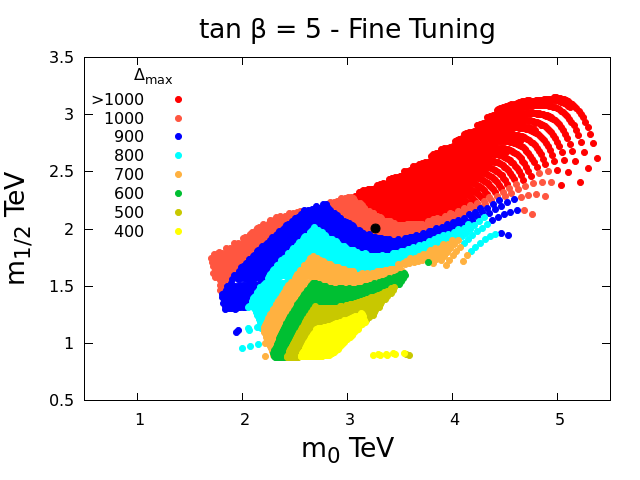}%
\end{minipage}\hfill{}%
\begin{minipage}[t][1\totalheight][c]{0.45\columnwidth}%
\includegraphics[scale=0.32]{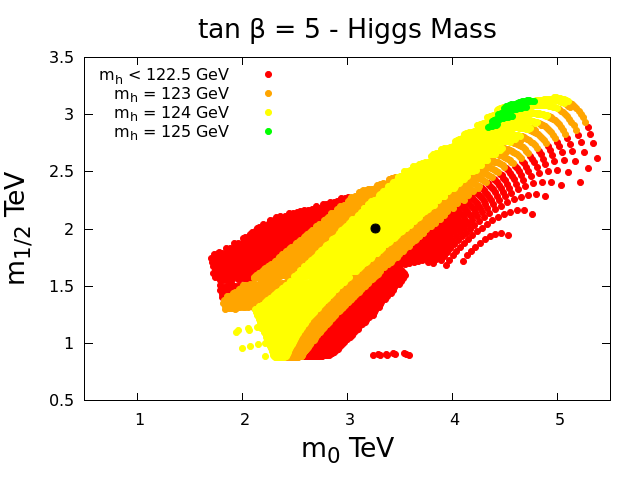}%
\end{minipage}

\caption{In the left panel, the fine tuning in the parameter space is shown in the $m_0-m_{1/2}$ plane,
while the the right panel shows the values of $m_h$. All with a fixed value of $M_{Z'} \approx 3.8$ TeV, and $\tan\beta = 5$.} 

\label{fig:tb5}
\end{figure}

\begin{figure}[H]
\begin{minipage}[t][1\totalheight][c]{0.45\columnwidth}%
\includegraphics[scale=0.32]{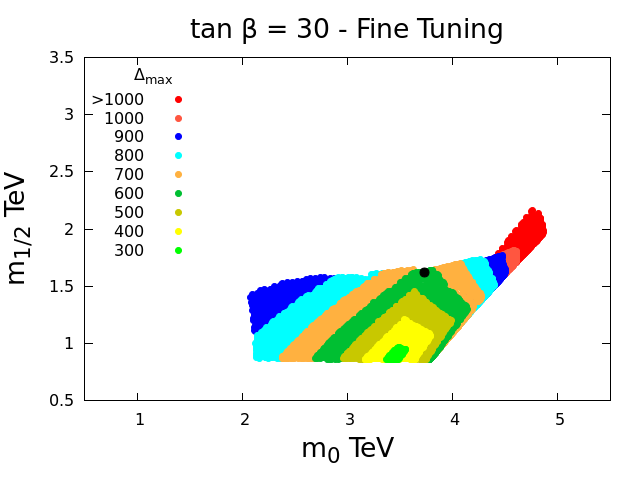}%
\end{minipage}\hfill{}%
\begin{minipage}[t][1\totalheight][c]{0.45\columnwidth}%
\includegraphics[scale=0.32]{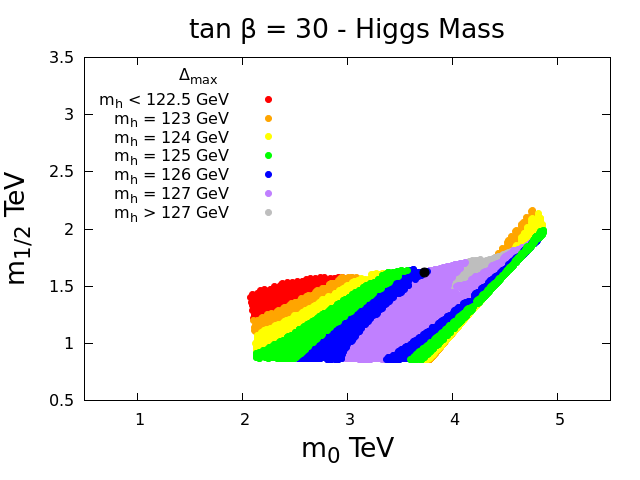}%
\end{minipage}

\caption{In the left panel, the fine tuning in the parameter space is shown in the $m_0-m_{1/2}$ plane,
while the the right panel shows the values of $m_h$. All with a fixed value of $M_{Z'} \approx 3.8$ TeV, and $\tan\beta = 30$.}

\label{fig:tb30}
\end{figure}

Scanning over values of $\lambda_3(\text{GUT}) \sim \{-3,0\},$ $\kappa_{1,2,3}(\text{GUT})\sim\{0,3\}$,
taking specific values of $\tan \beta = 5$ and $30$, while fixing
$s = 10$ TeV (i.e. $M_{Z'} \approx 3.8$ TeV). The cuts we applied
are rather conservative (see \citep{Athron:2013ipa}) as we require
a gluino mass $m_{\tilde{g}} > 1.4$ TeV. The Higgs mass is required
to be within the range $123 < m_h < 127$ GeV.

From the right panel in Fig.~\ref{fig:tb5}, one can see that small
values of $\tan \beta$ can hardly produce a Higgs mass larger than 124 GeV.
From our results we notice that the gluino mass can be large ($> 1.5$
TeV) in that region, however, fine tuning becomes larger as we approach
lower values of $\tan \beta$ (c.f. Fig.~\ref{fig:tb30}). The benchmark
point (appears as a black dot in the Figures) for the $\tan \beta = 5$
case corresponds to a Higgs mass of 124 GeV, $m_{\tilde{g}}\sim 1.7$
TeV, and fine tuning $\Delta \sim 1000$, which is $\sim 0.1\%$ tuning.

One the other hand, in regions where $\tan \beta$ is large, as in
Fig.~\ref{fig:tb30}, it is easy to find a Higgs mass above 125 GeV and
fine tuning is slightly lowered. However, as one approaches larger
and larger values it becomes somewhat difficult to find a gluino mass
larger than $1.5$ TeV. Therefore, moderate values of $\tan \beta$
are favored in this model from phenomenological and naturalness standpoints.
The benchmark point for $\tan \beta = 30$ corresponds to $m_h \sim 126.4$
GeV, $m_{\tilde{g}} \sim 1.4$ TeV, and fine tuning in the $\sim 0.2\%$
level ($\Delta \sim 600$), which is slightly better than the previous
case.

\section{Conclusions}

We have investigated regions of the parameter space of the cE$_6$SSM
where $\tan \beta$ is as low as 5 and as high as 30. Moreover, we
took into account the latest experimental limits on SUSY particles
as well as the measured value of the Higgs boson. We find that, in
general, fine tuning in the Electroweak sector lies in a
level between $0.2\%-0.1\%$. Small $\tan \beta$ regions are characterized
by Higgs mass between 123 and 125 GeV, and a gluino mass that can be larger
than 1.5 TeV. The fine tuning is more severe in this region
of the parameter space. On the other hand, large $\tan \beta$ regions
are associated with larger Higgs mass ranging from 123 to 127 GeV, 
but gluino mass that tend to be smaller than 1.5 TeV. However,
the fine tuning is slightly less than that in the very low $\tan \beta$
regime. We can then conclude that moderate values of $\tan \beta$
are favoured by naturalness in the cE$_6$SSM. Finally, in future studies,
one can study the effects of including radiative corrections (e.g. 
the one-loop Coleman-Weinberg potential) into the definition of fine tuning.

\section*{Acknowledgements}
I am grateful to my supervisor Stephen F. King and our collaborator
Peter Athron, for they made carrying out this work possible. I would like to thank 
the organising committee of the 7th Saudi Students Conference in the UK for putting together a 
wonderful meeting. The work of MB is funded by King Saud University (Riyadh, Saudi Arabia). 

\bibliographystyle{h-physrev}
\addcontentsline{toc}{section}{\refname}\bibliography{cE6SSMtb.NE.10}

\begin{thebibliography}{10}

\bibitem{'tHooft:1979bh}
G.~'t~Hooft,
\newblock NATO Adv.Study Inst.Ser.B Phys. {\bf 59}, 135 (1980).

\bibitem{PhysRevD.13.3333}
E.~Gildener and S.~Weinberg,
\newblock Phys. Rev. D {\bf 13}, 3333 (1976).

\bibitem{Susskind:1978ms}
L.~Susskind,
\newblock Phys.Rev. {\bf D20}, 2619 (1979).

\bibitem{Martin:1997ns}
S.~P. Martin,
\newblock (1997), hep-ph/9709356.

\bibitem{ATLAS-CONF-2013-053}
CERN Report No. ATLAS-CONF-2013-053, 2013 (unpublished).

\bibitem{ATLAS-CONF-2013-062}
CERN Report No. ATLAS-CONF-2013-062, 2013 (unpublished).

\bibitem{King:2005jy}
S.~King, S.~Moretti, and R.~Nevzorov,
\newblock Phys.Rev. {\bf D73}, 035009 (2006), hep-ph/0510419.

\bibitem{King:2005my}
S.~King, S.~Moretti, and R.~Nevzorov,
\newblock Phys.Lett. {\bf B634}, 278 (2006), hep-ph/0511256.

\bibitem{Hall:2012mx}
J.~P. Hall and S.~F. King,
\newblock JHEP {\bf 1301}, 076 (2013), 1209.4657.

\bibitem{Athron:2013ipa}
P.~Athron, M.~Binjonaid, and S.~F. King,
\newblock Phys.Rev. {\bf D87}, 115023 (2013), 1302.5291.

\bibitem{Athron:2009bs}
P.~Athron, S.~King, D.~Miller, S.~Moretti, and R.~Nevzorov,
\newblock Phys.Rev. {\bf D80}, 035009 (2009), 0904.2169.

\bibitem{Athron:2009ue}
P.~Athron {\em et~al.},
\newblock Phys.Lett. {\bf B681}, 448 (2009), 0901.1192.

\bibitem{Athron:2010zz}
P.~Athron {\em et~al.},
\newblock Nucl.Phys.Proc.Suppl. {\bf 200-202}, 120 (2010).

\bibitem{Athron:2011wu}
P.~Athron, S.~King, D.~Miller, S.~Moretti, and R.~Nevzorov,
\newblock Phys.Rev. {\bf D84}, 055006 (2011), 1102.4363.

\bibitem{Athron:2012pw}
P.~Athron, D.~Stockinger, and A.~Voigt,
\newblock Phys.Rev. {\bf D86}, 095012 (2012), 1209.1470.

\bibitem{Athron:2012sq}
P.~Athron, S.~King, D.~Miller, S.~Moretti, and R.~Nevzorov,
\newblock Phys.Rev. {\bf D86}, 095003 (2012), 1206.5028.

\bibitem{Hall:2009aj}
J.~P. Hall and S.~F. King,
\newblock JHEP {\bf 0908}, 088 (2009), 0905.2696.

\bibitem{Hall:2010ix}
J.~Hall {\em et~al.},
\newblock Phys.Rev. {\bf D83}, 075013 (2011), 1012.5114.

\bibitem{Hall:2011zq}
J.~P. Hall and S.~F. King,
\newblock JHEP {\bf 1106}, 006 (2011), 1104.2259.

\bibitem{Belyaev:2012jz}
A.~Belyaev, J.~P. Hall, S.~F. King, and P.~Svantesson,
\newblock Phys.Rev. {\bf D87}, 035019 (2013), 1211.1962.

\bibitem{Belyaev:2012si}
A.~Belyaev, J.~P. Hall, S.~F. King, and P.~Svantesson,
\newblock Phys.Rev. {\bf D86}, 031702 (2012), 1203.2495.

\bibitem{Howl:2008xz}
R.~Howl and S.~King,
\newblock JHEP {\bf 0805}, 008 (2008), 0802.1909.

\bibitem{Howl:2009ds}
R.~Howl and S.~King,
\newblock Phys.Lett. {\bf B687}, 355 (2010), 0908.2067.

\bibitem{King:2007uj}
S.~King, S.~Moretti, and R.~Nevzorov,
\newblock Phys.Lett. {\bf B650}, 57 (2007), hep-ph/0701064.

\bibitem{Callaghan:2012rv}
J.~C. Callaghan and S.~F. King,
\newblock JHEP {\bf 1304}, 034 (2013), 1210.6913.

\bibitem{Callaghan:2013kaa}
J.~C. Callaghan, S.~F. King, and G.~K. Leontaris,
\newblock (2013), 1307.4593.

\bibitem{Ellis:1986yg}
J.~R. Ellis, K.~Enqvist, D.~V. Nanopoulos, and F.~Zwirner,
\newblock Mod.Phys.Lett. {\bf A1}, 57 (1986).

\bibitem{Barbieri:1987fn}
R.~Barbieri and G.~Giudice,
\newblock Nucl.Phys. {\bf B306}, 63 (1988).

\bibitem{Chankowski:1997zh}
P.~H. Chankowski, J.~R. Ellis, and S.~Pokorski,
\newblock Phys.Lett. {\bf B423}, 327 (1998), hep-ph/9712234.

\bibitem{Agashe:1997kn}
K.~Agashe and M.~Graesser,
\newblock Nucl.Phys. {\bf B507}, 3 (1997), hep-ph/9704206.

\bibitem{Wright:1998mk}
D.~Wright,
\newblock (1998), hep-ph/9801449.

\bibitem{Kane:1998im}
G.~L. Kane and S.~King,
\newblock Phys.Lett. {\bf B451}, 113 (1999), hep-ph/9810374.

\bibitem{BasteroGil:1999gu}
M.~Bastero-Gil, G.~L. Kane, and S.~King,
\newblock Phys.Lett. {\bf B474}, 103 (2000), hep-ph/9910506.

\bibitem{Kang:2012sy}
Z.~Kang, J.~Li, and T.~Li,
\newblock JHEP {\bf 1211}, 024 (2012), 1201.5305.

\bibitem{Antusch:2012gv}
S.~Antusch, L.~Calibbi, V.~Maurer, M.~Monaco, and M.~Spinrath,
\newblock JHEP {\bf 01}, 187 (2013), 1207.7236.

\bibitem{Perelstein:2012qg}
M.~Perelstein and B.~Shakya,
\newblock (2012), 1208.0833.

\end{thebibliography}
 
\end{document}